\documentstyle[psfig]{elsart}
\begin{document}
\begin{frontmatter}
\title{The anomalous process {\boldmath $\gamma\pi\rightarrow\pi\pi$}
to two loops}
\author{Torben Hannah\thanksref{email}}
\thanks[email]{Present e-mail address: torben@edlund.dk}
\address{Nordita, Blegdamsvej 17, DK-2100 Copenhagen {\O}, Denmark}
\begin{abstract}
The amplitude for the anomalous process $\gamma\pi\rightarrow\pi\pi$
is evaluated to two loops in the chiral expansion by means of a
dispersive method. The two new coupling constants that enter at
this order are estimated via sum rules derived from a non-perturbative
chiral approach. With these coupling constants fixed, the numerical
results are given and compared with the available experimental
information.
\end{abstract}
\end{frontmatter}

\section{Introduction}
\label{sec:Intro}

At low energies, the strong interaction between pions can be described
by the effective field theory of QCD called chiral perturbation theory
(ChPT) \cite{ref:We79,ref:GL84,ref:GL85,ref:Leut94}. In this effective
field theory the interaction between pions is analyzed in terms
of a systematic expansion in powers of the low external momenta and
the small pion mass. This chiral expansion is equivalent to an
expansion in the number of loops and works very well for processes
involving pions \cite{ref:RevChPT}.

In the normal intrinsic parity sector, the chiral expansion starts
at $O(p^2)$ and is obtained from the leading order chiral Lagrangian.
In this sector the one-loop or $O(p^4)$ corrections were
extensively treated in the works by Gasser and Leutwyler
\cite{ref:GL84,ref:GL85}, and many processes have now been calculated
to this order. The further extension to two loops of $O(p^6)$ has
also been undertaken with several two-flavour calculations already
published in the normal sector 
\cite{ref:GM91,ref:BCT98,ref:Knecht95,ref:Bijetal96,ref:BGS94}. 

For the abnormal intrinsic parity sector, the chiral expansion starts
at $O(p^4)$ and is obtained from the Wess-Zumino-Witten effective
action \cite{ref:WZ71}. In this sector the one-loop corrections of
$O(p^6)$ have also been analyzed \cite{ref:BBC90}, and some
anomalous processes have been calculated to this order \cite{ref:Bij93}.
However, contrary to the case in the normal sector, no two-loop
calculations of $O(p^8)$ have been published for processes in the
anomalous sector.

In this paper the anomalous process $\gamma\pi\rightarrow\pi\pi$
is calculated to two loops using a dispersive method. This process
is important for the theory of chiral anomalies and has previously
been calculated to both leading \cite{ref:Adler71} and one-loop
\cite{ref:BBC90a} order in the chiral expansion. However, both of
these results are somewhat below the present experimental data
\cite{ref:Antipov97}. This has motivated new and more precise
experiments, which will be performed at different facilities such
as CERN \cite{ref:COMPASS}, FNAL \cite{ref:SELEX}, and CEBAF
\cite{ref:CEBAF}.

It is therefore important to evaluate the two-loop corrections and
compare the result with the experimental information. This is the
purpose of the present paper, which is organized as follows.
In Sect. \ref{sec:Kin}, the notation and kinematics of the process
$\gamma\pi\rightarrow\pi\pi$ are given, whereas previous ChPT results
are reviewed in Sect. \ref{sec:ChPT}. The general dispersive formula
is derived in Sect. \ref{sec:Disp}, together with the explicit
calculation of the two-loop corrections. In Sect. \ref{sec:Nonper},
this two-loop result is used in a non-perturbative chiral approach
in order to estimate the two new coupling constants from sum rules.
The numerical results are given in Sect. \ref{sec:Num}, including the
comparison with experiments, and Sect. \ref{sec:Con} consists of a
short conclusion.

\section{Kinematics}
\label{sec:Kin}

The amplitude for the anomalous process
\begin{equation}
\label{eq:kingppp}
\gamma (q)\pi^-(p_1)\rightarrow\pi^-(p_2)\pi^0(p_0)
\end{equation}
is given in terms of the scalar function $F_{3\pi}(s,t,u)$ as
\begin{equation}
\label{eq:gamp}
A = iF_{3\pi}(s,t,u)\epsilon^{\mu\nu\alpha\beta}
\epsilon_{\mu}p_{1\nu}p_{2\alpha}p_{0\beta} ,
\end{equation}
where $s=(p_1+q)^2$, $t=(p_1-p_2)^2$, and $u=(p_1-p_0)^2$ are
the Mandelstam variables. In the following it is assumed that
all particles are real. In this case $s+t+u=3M^2_{\pi}$, so
that in the center of mass system one has
\begin{eqnarray}
\label{eq:tucm}
t & = & \frac{1}{2}\left[ 3M^2_{\pi}-s+(s-M^2_{\pi})
\sqrt{1-4M^2_{\pi}/s}\cos\theta \right] , \nonumber \\
u & = & \frac{1}{2}\left[3M^2_{\pi}-s-(s-M^2_{\pi})
\sqrt{1-4M^2_{\pi}/s}\cos\theta \right]
\end{eqnarray}
with $\theta$ the center of mass scattering angle. Because of
isospin symmetry, the other $\gamma\pi\rightarrow\pi\pi$ reactions
are given by the same scalar function $F_{3\pi}(s,t,u)$, which is
fully symmetrical in its arguments. The total cross section is obtained
from the expression
\begin{equation}
\label{eq:totalcross}
\sigma (s) = \frac{1}{1024\pi}\frac{(s-4M_{\pi}^2)^{3/2}}{s^{1/2}}
(s-M_{\pi}^2)\int_0^{\pi}d\theta\sin^3\theta |F_{3\pi}(s,t,u)|^2 ,
\end{equation}
and the partial wave expansion of the scalar function $F_{3\pi}(s,t,u)$
is of the form
\begin{equation}
\label{eq:expar}
F_{3\pi}(s,t,u) = \sum_{odd\;l}f_l(s)P_l^{\prime}(\cos\theta ) .
\end{equation}
The projection onto the lowest $P$ partial wave is given by
\begin{equation}
\label{eq:pwave}
f_1(s) = \frac{3}{8\pi}\int d\Omega\sin^2\theta F_{3\pi}(s,t,u) ,
\end{equation}
and the higher partial waves can be projected out in similar ways.
This partial wave expansion contains only odd $l$ since the pions in
the final state have the total isospin $I=1$. In the unitarity relation,
the complete set of intermediate states are the same as for $\pi\pi$
scattering. Within the elastic approximation containing only the
two-pion intermediate state, the unitarity relation is
\begin{equation}
\label{eq:funi}
{\rm Im}f_l(s) = \sigma (s)f_l^{\ast}(s)t^1_l(s)
\end{equation}
with $\sigma (s)=\sqrt{(s-4M_{\pi}^2)/s}$ the phase-space factor 
and $t^1_l$ the $\pi\pi$ partial wave with isospin $I=1$ and
angular momentum $l$. This unitarity relation implies that the
phase of $f_l$ will coincide with the $\pi\pi$ phase shift
$\delta^1_l$ in accordance with Watson's final-state theorem
\cite{ref:Watson54}. Inelasticities due to other intermediate
states such as the four-pion or the $K\bar{K}$ states remain very
small below 1 GeV and are completely negligible at low energies.

\section{Chiral expansion}
\label{sec:ChPT}

Anomalous processes start at $O(p^4)$ in the chiral expansion
and are obtained from the Wess-Zumino-Witten effective action
at tree level. With only the electromagnetic interaction as
external fields, these anomalous processes contain exclusively
an odd number of pseudoscalars, i.e., they have abnormal or odd
intrinsic parity. For the process $\gamma\pi\rightarrow\pi\pi$,
the leading order chiral result for the scalar function
$F_{3\pi}(s,t,u)$ is given by \cite{ref:WZ71,ref:Adler71}
\begin{equation}
\label{eq:ChPT0}
F_{3\pi}(s,t,u) = F_{3\pi}^{(0)} = \frac{eN_c}{12\pi^2F_{\pi}^3} ,
\end{equation}
where $N_c=3$ is the number of colors and $F_{\pi}=92.4$ MeV is
the pion decay constant. Another related anomalous process is the
decay $\pi^0\rightarrow\gamma\gamma$, where the $F_{\pi\gamma\gamma}$
coupling obtained at leading order is of the form
\cite{ref:WZ71,ref:Adler69}
\begin{equation}
\label{eq:pigg}
F^{(0)}_{\pi\gamma\gamma} = \frac{\alpha N_c}{3\pi F_{\pi}} .
\end{equation}
These two results are exact in the soft pion and soft photon limit
where they are related to each other by the low-energy theorem
$F_{3\pi}=F_{\pi\gamma\gamma}/eF_{\pi}^2$. For the decay
$\pi^0\rightarrow\gamma\gamma$, the prediction from Eq.
(\ref{eq:pigg}) is in excellent agreement with the experimental
information, thus providing important evidence for the
three-color nature of the strong interaction. However, for
the process $\gamma\pi\rightarrow\pi\pi$, the corresponding
prediction $F_{3\pi}=9.7$ ${\rm GeV}^{-3}$ is somewhat lower than
the experimental measurement $F_{3\pi}=12.9\pm 0.9\pm 0.5$
${\rm GeV}^{-3}$ \cite{ref:Antipov97}, which could even be said to
favor the value $N_c=4$.

It is therefore of importance to calculate the corrections to
the soft pion and soft photon limit. Within ChPT these corrections
can indeed be calculated in a systematic manner. The first correction
is the one-loop contribution of $O(p^6)$, which has previously been
calculated for both processes. In the case of the decay
$\pi^0\rightarrow\gamma\gamma$, these one-loop corrections turn
out to be very small \cite{ref:DHL85}, i.e., they do not
spoil the excellent agreement with the experimental information.
On the other hand, for the process $\gamma\pi\rightarrow\pi\pi$,
the one-loop contributions are larger and they will therefore be
of importance when comparing with the experimental measurement. The
expression for the scalar function $F_{3\pi}(s,t,u)$ to this
order in the chiral expansion is \cite{ref:BBC90a}
\begin{equation}
\label{eq:ChPT1}
F_{3\pi}(s,t,u) = F_{3\pi}^{(0)}\left[ 1-\frac{64\pi^2}{3e}
C_2^r(\mu )(s+t+u)+C_l^{\pi} \right] ,
\end{equation}
where the term $C_l^{\pi}$ contains the contributions
from the loops and is given by
\begin{eqnarray}
\label{eq:Cloops}
C_l^{\pi} & = & \frac{1}{96\pi^2F_{\pi}^2}\left[
-(s+t+u)\log\frac{M_{\pi}^2}{\mu^2}+\frac{5}{3}(s+t+u) \right]
\nonumber \\
&& +\frac{1}{24\pi^2F_{\pi}^2}\left[ F(s)+F(t)+F(u) \right] 
\end{eqnarray}
with $F(x)$ expressed in terms of the standard one-loop function
$\bar{J}(x)$ as
\begin{equation}
\label{eq:Fexp}
F(x) = 4\pi^2(x-4M_{\pi}^2)\bar{J}(x)-\mbox{$\frac{1}{2}$}x .
\end{equation}
The expression (\ref{eq:ChPT1}) converges to the chiral anomaly
$F_{3\pi}^{(0)}$ in the soft pion and soft photon limit and
is given in terms of the renormalized low-energy constant
$C_2^r$ from the $O(p^6)$ anomalous chiral
Lagrangian \cite{ref:BBC90}. This low-energy constant depends
on the renormalization scale $\mu$ and is needed to absorb
divergences in the one-loop calculation. In principle, $C_2^r$
should be determined phenomenologically, preferably
from other observables, but at present this appears to
be rather out of reach. Therefore, this low-energy constant
has been estimated using the assumption of vector resonance
saturation, which is known to work well for the $O(p^4)$
non-anomalous low-energy constants \cite{ref:Ecker89}. Assuming
that the same is the case for the $O(p^6)$ anomalous low-energy
constants, one has \cite{ref:BBC90a}
\begin{equation}
\label{eq:cr2}
C_2^r(\mu ) = -\frac{3e}{128\pi^2M_{\rho}^2}
\end{equation}
with the renormalization scale typical chosen at the resonance
scale $\mu =M_{\rho}=770$ MeV. Having fixed the value of this
low-energy constant, it is possible to obtain the prediction for
the one-loop expression (\ref{eq:ChPT1}). This improves the
agreement with the experimental measurement \cite{ref:Antipov97}.
However, the prediction is still on the lower side of the data
\cite{ref:BBC90a,ref:COMPASS,ref:SELEX}.

Therefore, it is important with both theoretical and experimental
improvements. Indeed, significant improvements on the experimental
side are expected from several new experiments
\cite{ref:COMPASS,ref:SELEX,ref:CEBAF}. A similar improvement
on the theore\-tical side would involve the calculation
of the two-loop corrections. This could be done by a full
field theory calculation to two loops. However, these two-loop
corrections can also be calculated by a dispersive method,
which will be the subject of the next section.

\section{Dispersive representation}
\label{sec:Disp}

\subsection{Derivation of the dispersive formula}
\label{subsec:fispfor}

The anomalous process $\gamma\pi\rightarrow\pi\pi$ is in many
ways similar to the $\pi^0\pi^0\rightarrow\pi^0\pi^0$ scattering
process. They are both described in terms of a single scalar
function, which is fully symmetrical in the $s$, $t$, and $u$
variables. The $\pi\pi$ scattering process can be described by
Roy equations \cite{ref:Roy71} derived from the fundamental
principles of analyticity, crossing, and unitarity. When these Roy
equations are combined with the chiral expansion, the general
structure of the $\pi\pi$ scattering amplitude can be obtained
from a dispersive representation to two loops in the chiral expansion
\cite{ref:Knecht95,ref:SSF93}.

The same is also the case for the process $\gamma\pi\rightarrow\pi\pi$.
In order to show this, one starts with a fixed-t dispersion relation
for the scalar function $F_{3\pi}(s,t,u)$. Fixed-t dispersion relations
with two subtractions are known to exist for $\pi\pi$ scattering due
to the Froissart bound. Assuming that the same is also the case
for the process $\gamma\pi\rightarrow\pi\pi$, one has the fixed-t
dispersion relation
\begin{equation}
\label{eq:fixedt}
F_{3\pi}(s,t) = C(t) + \frac{1}{\pi}\int_{4M_{\pi}^2}^{\infty}ds'
\frac{{\rm Im}F_{3\pi}(s',t)}{s'^2}\left( \frac{s^2}{s'-s}+
\frac{u^2}{s'-u}\right) ,
\end{equation}
where $u=3M_{\pi}^2-s-t$. The $s\leftrightarrow u$ symmetry of
$F_{3\pi}(s,t,u)$ implies that there is no subtraction constant
linear in $s$. Therefore, the only subtraction constant is $C(t)$,
where the $t$ dependence can be obtained from the
$s\leftrightarrow t$ symmetry $F_{3\pi}(0,t)=F_{3\pi}(t,0)$.
This implies that $C(t)$ can be written as
\begin{eqnarray}
\label{eq:Ct}
C(t) & = & C(0)+\frac{1}{\pi}\int_{4M_{\pi}^2}^{\infty}ds'
\frac{{\rm Im}F_{3\pi}(s',0)}{s'^2}\left( \frac{t^2}{s'-t}+
\frac{(3M_{\pi}^2-t)^2}{s'-3M_{\pi}^2+t} \right) \nonumber \\
&& -\frac{1}{\pi}\int_{4M_{\pi}^2}^{\infty}ds'\frac{{\rm Im}
F_{3\pi}(s',t)}{s'^2}\frac{(3M_{\pi}^2-t)^2}{s'-3M_{\pi}^2+t} ,
\end{eqnarray}
which can be inserted in Eq. (\ref{eq:fixedt}). By construction,
the resulting dispersion relation exhibits $s\leftrightarrow u$
symmetry for fixed $t$, whereas $s\leftrightarrow t$ symmetry
for fixed $u$ is not manifest. In order to impose this latter
symmetry one can expand the absorptive part of $F_{3\pi}(s',t)$
in partial waves using Eq. (\ref{eq:expar}). Writing this expansion
in the form
\begin{equation}
\label{eq:ImF}
{\rm Im}F_{3\pi}(s',t) = {\rm Im}f_1(s') + {\rm Im}\Phi (s',t) ,
\end{equation}
the higher partial waves with $l\geq 3$ are contained in
the function ${\rm Im}\Phi (s',t)$, which is given by
\begin{equation}
\label{eq:ImPhi}
{\rm Im}\Phi (s',t) = \sum_{l\geq 3}{\rm Im}f_l(s')
P_l^{\prime}(\cos\theta ) .
\end{equation}
With this decomposition of the partial waves, it is possible to
rewrite Eq. (\ref{eq:fixedt}) with $C(t)$ given by Eq.
(\ref{eq:Ct}) as
\begin{equation}
\label{eq:Ffinal}
F_{3\pi}(s,t,u) = \hat{F}_{3\pi}(s,t,u) + \Phi_{3\pi}(s,t,u) ,
\end{equation}
where $\hat{F}_{3\pi}(s,t,u)$ and $\Phi_{3\pi}(s,t,u)$ are
given by
\begin{eqnarray}
\label{eq:Fhat}
\hat{F}_{3\pi}(s,t,u) & = & C(0) +\frac{1}{\pi}
\int_{4M_{\pi}^2}^{\infty}ds'
\frac{{\rm Im}f_1(s')}{s'^2}\left( \frac{s^2}{s'-s}+\frac{t^2}{s'-t}
+\frac{u^2}{s'-u}\right) , \nonumber \\
\Phi_{3\pi}(s,t,u) & = & \frac{1}{\pi}\int_{4M_{\pi}^2}^{\infty}ds'
\frac{{\rm Im}\Phi (s',t)}{s'^2}\left( \frac{s^2}{s'-s}
+\frac{u^2}{s'-u}-\frac{(3M_{\pi}^2-t)^2}{s'-3M_{\pi}^2+t}
\right) \nonumber \\
&& + \frac{1}{\pi}\int_{4M_{\pi}^2}^{\infty}ds'
\frac{{\rm Im}\Phi (s',0)}{s'^2}\left( \frac{t^2}{s'-t}+
\frac{(3M_{\pi}^2-t)^2}{s'-3M_{\pi}^2+t} \right) .
\end{eqnarray}
This is the Roy equation for the anomalous process
$\gamma\pi\rightarrow\pi\pi$. One observes that $\hat{F}_{3\pi}(s,t,u)$
is now fully symmetrical in $s$, $t$, and $u$, whereas this
symmetry is not manifest in $\Phi_{3\pi}(s,t,u)$. However, at low
energies, the absorptive part of the higher partial waves
with $l\geq 3$ is negligible. This implies that in practice
$\Phi_{3\pi}(s,t,u)$ can be treated in a simple way as a small,
real correction at low energies. This fact makes
the corresponding Roy equations for $\pi\pi$ scattering
very useful.

In ChPT the absorptive part of the higher partial waves is indeed
suppressed. Within this methodology, these higher partial waves only
start at $O(p^6)$ in the chiral expansion. Since the corresponding
$\pi\pi$ partial waves start at $O(p^4)$, perturbative unitarity
implies that the absorptive part of the higher partial waves
with $l\geq 3$ is of $O(p^{10})$ or higher in the chiral expansion.
Thus, up to this order, the term $\Phi_{3\pi}(s,t,u)$ in Eq.
(\ref{eq:Ffinal}) vanishes, and the full amplitude is given entirely
by the function $\hat{F}_{3\pi}(s,t,u)$. In order for the full
amplitude to formally satisfy the chiral anomaly in the soft pion
and soft photon limit, one may write the subtraction constant as
$C(0)=F_{3\pi}^{(0)}[1+\bar{C}(s+t+u)]$, where $s+t+u=3M_{\pi}^2$.
Thus, assuming for the moment that two subtractions
are indeed sufficient, the full amplitude to $O(p^8)$
in the chiral expansion may be written as
\begin{eqnarray}
\label{eq:Ffull}
F_{3\pi}(s,t,u) & = & F_{3\pi}^{(0)}\left[1+\bar{C}(s+t+u)\right]
\nonumber \\
&& +\frac{1}{\pi}\int_{4M_{\pi}^2}^{\infty}ds'
\frac{{\rm Im}f_1(s')}{s'^2}\left( \frac{s^2}{s'-s}+\frac{t^2}{s'-t}
+\frac{u^2}{s'-u}\right) .
\end{eqnarray}
In this dispersive representation, the absorptive part of the lowest
partial wave can be determined from unitarity. In the general
unitarity relation, the $2n$-pion invariant phase space is of
$O(p^{4n-4})$, the amplitude for multi-pion scattering is
dominantly of $O(p^2)$, and the amplitude for multi-pion
photo-production is at least of $O(p^4)$. Consequently, intermediate
states containing more than two pions are suppressed at least up to
$O(p^{10})$ in the chiral expansion. Therefore, within SU(2) ChPT,
the absorptive part of the lowest partial wave is given by elastic
unitarity to $O(p^8)$ as
\begin{eqnarray}
\label{eq:puni}
{\rm Im}f_1^{(0)}(s) & = & 0 , \nonumber \\
{\rm Im}f_1^{(1)}(s) & = & \sigma (s)f_1^{(0)}(s)t^{1(0)}_1(s) ,
\nonumber \\
{\rm Im}f_1^{(2)}(s) & = & \sigma (s)\left[ f_1^{(0)}(s){\rm Re}
t^{1(1)}_1(s)+{\rm Re}f_1^{(1)}(s)t^{1(0)}_1(s) \right] ,
\end{eqnarray}
where $f_1^{(n)}$ is the $P$ partial wave of $O(p^{2n+4})$ and
$t_1^{1(n)}$ the corresponding $\pi\pi$ partial wave of $O(p^{2n+2})$
\cite{ref:GL84}. In SU(3) ChPT, one must also include the
inelasticity from the $K\bar{K}$ intermediate state. However,
since this inelasticity is completely negligible at low energies,
only SU(2) ChPT will be considered in the following.

\subsection{The amplitude to two loops}
\label{subsec:Two}

From the dispersive representation (\ref{eq:Ffull}), one
obtains straightforwardly the one-loop ChPT formula (\ref{eq:ChPT1})
by setting ${\rm Im}f_1={\rm Im}f_1^{(1)}$. In this case
the subtraction constant $\bar{C}$ can be expressed in terms of the 
low-energy constant $C_2^r$ and chiral logarithms. However, in order
to calculate the amplitude to two loops, it is necessary to use one
subtraction more in the dispersive representation. This is due to the
fact that the absorptive part of $f_1^{(2)}$ behaves as $s^2$ modulo
log factors. Thus, the amplitude to two loops can be obtained from
the following dispersion relation:
\begin{eqnarray}
\label{eq:Ffull1}
F_{3\pi}(s,t,u) & = & F_{3\pi}^{(0)}\left[1+\bar{C}(s+t+u)+\bar{D}
(s^2+t^2+u^2) \right] \nonumber \\
&& +\frac{1}{\pi}\int_{4M_{\pi}^2}^{\infty}ds'
\frac{{\rm Im}f_1(s')}{s'^3}\left( \frac{s^3}{s'-s}+\frac{t^3}{s'-t}
+\frac{u^3}{s'-u}\right) ,
\end{eqnarray}
where $\bar{C}$ and $\bar{D}$ are the two subtraction constants and
${\rm Im}f_1$ is given by
${\rm Im}f_1={\rm Im}f_1^{(1)}+{\rm Im}f_1^{(2)}$ with
${\rm Im}f_1^{(1)}$ and ${\rm Im}f_1^{(2)}$ determined from Eq.
(\ref{eq:puni}). This dispersion relation can be evaluated with
the same methodology as has been applied in the calculation of
the $\pi\pi$ scattering amplitude \cite{ref:Knecht95} and the pion
form factors \cite{ref:GM91} to two loops. With the use of this
methodology, the result can be written as
\begin{eqnarray}
\label{eq:disp2}
F_{3\pi}(s,t,u) & = & F_{3\pi}^{(0)}\left[ 1+\bar{C}(s+t+u)+
\bar{D}(s^2+t^2+u^2) \right. \nonumber \\
&& \left. +U^A(s)+U^A(t)+U^A(u)+U^{\Delta}(s)+U^{\Delta}(t)+
U^{\Delta}(u) \right] ,
\end{eqnarray}
where the $U^A$ term can be expressed in the compact analytic form:
\begin{eqnarray}
\label{eq:uni2}
U^A(x) & = & \frac{M_{\pi}^2}{16\pi^2F_{\pi}^2}\left\{
\frac{x}{9M_{\pi}^2}\left[ 1+24\pi^2\sigma^2(x)\bar{J}(x)\right]
-\frac{x^2}{60M_{\pi}^4} \right\}
+\left( \frac{M_{\pi}^2}{16\pi^2F_{\pi}^2}\right)^2
\nonumber \\
&& \times\left\{ \left[\bar{l}_2-\bar{l}_1+\frac{6M_{\pi}^2\bar{l}_4}{x}
+\frac{9M_{\pi}^2\bar{c}_1}{x}\right]\frac{x^2}{27M_{\pi}^4}
\left[ 1+24\pi^2\sigma^2(x)\bar{J}(x)\right] \right. \nonumber \\
&& -\frac{x^2}{30M_{\pi}^4}\bar{l}_4-\frac{x^2}{20M_{\pi}^4}\bar{c}_1
+\frac{3191}{6480}\frac{x^2}{M_{\pi}^4}+\frac{223}{216}
\frac{x}{M_{\pi}^2}-\frac{16}{9} \nonumber \\
&& -\frac{\pi^2x}{540M_{\pi}^2}\left( 37\frac{x}{M_{\pi}^2}+15\right)
+\frac{4\pi^2}{27}\left( 7\frac{x^2}{M_{\pi}^4}-151\frac{x}{M_{\pi}^2}
+99\right)\bar{J}(x) \nonumber \\
&& +\frac{2\pi^2M_{\pi}^2}{9x}\left(\frac{x^3}{M_{\pi}^6}
-30\frac{x^2}{M_{\pi}^4}+78\frac{x}{M_{\pi}^2}-128\right)\bar{K}_1(x)
\nonumber \\
&& \left. +8\pi^2\left(\frac{x^2}{M_{\pi}^4}-\frac{13}{3}
\frac{x}{M_{\pi}^2}-2\right)\bar{K}_4(x) \right\} .
\end{eqnarray}
The explicit expressions for the functions $\bar{K}_1(x)$ and
$\bar{K}_4(x)$ are given in Ref. \cite{ref:Knecht95}, where
these functions were introduced in the evaluation of the
$\pi\pi$ scattering amplitude to two loops. They are analytic
functions with cuts starting at the $\pi\pi$ threshold and can
be expressed in terms of the standard one-loop function
$\bar{J}(x)$. As for the other term $U^{\Delta}$, this part
does not have a compact analytic representation in terms of
elementary functions. However, it can be obtained numerically
from the dispersive representation:
\begin{equation}
\label{eq:ddelta}
U^{\Delta}(x) = \frac{x^3}{\pi}\int_{4M_{\pi}^2}^{\infty}
ds'\frac{\sigma (s'){\rm Re}f_1^{\Delta}(s')t_1^{1(0)}(s')}
{s'^3(s'-x)} .
\end{equation}
In this dispersive formula $t_1^{1(0)}$ is the lowest order
$\pi\pi$ $P$ partial wave and $f_1^{\Delta}$ is given by
\begin{equation}
\label{eq:fdelta}
f_1^{\Delta}(s) = \frac{3}{8\pi}\int d\Omega\sin^2\theta
\frac{1}{96\pi^2F_{\pi}^2}\left\{\frac{5}{3}(t+u)
+4[F(t)+F(u)]\right\}
\end{equation}
with $t$ and $u$ determined from Eq. (\ref{eq:tucm}) and $F(x)$ given
by Eq. (\ref{eq:Fexp}). The size of the corrections from
$U^{\Delta}$ are very small at low energies and they can therefore
in practice be neglected at these energies.

In the dispersive representation, Eq. (\ref{eq:disp2}), the
full amplitude to two loops is determined up to the subtraction
constants $\bar{C}$ and $\bar{D}$. These subtraction
constants may be parameterized in terms of the coupling constants
$\bar{c}_1$, $\bar{c}_2$, and $\bar{d}_2$ as
\begin{eqnarray}
\label{eq:cdbar}
\bar{C} & = & \frac{1}{16\pi^2F_{\pi}^2}\left[ (\bar{c}_1-
\mbox{$\frac{1}{6}$} )+\frac{M_{\pi}^2}{16\pi^2F_{\pi}^2}\bar{c}_2
\right] ,\nonumber \\
\bar{D} & = & \frac{1}{16\pi^2F_{\pi}^2}\left[ \frac{1}{60M_{\pi}^2}
+\frac{1}{16\pi^2F_{\pi}^2}\bar{d}_2\right] ,
\end{eqnarray}
where the coupling constant $\bar{c}_1$ can be expressed in terms of
the $O(p^6)$ anomalous low-energy constant $C_2^r$ and chiral
logarithms:
\begin{equation}
\label{eq:c1bar}
\bar{c}_1 = 16\pi^2F_{\pi}^2\left[ -\frac{64\pi^2}{3e}C_2^r(\mu )
-\frac{1}{96\pi^2F_{\pi}^2}\log\frac{M_{\pi}^2}{\mu^2}\right] .
\end{equation}
The coupling constants $\bar{c}_2$ and $\bar{d}_2$ enter at two-loop
order in the chiral expansion and contain contributions both from
two-loop diagrams and from the unknown renormalized low-energy
constants, which parameterize the $O(p^8)$ anomalous chiral Lagrangian.

\section{Non-perturbative chiral approach}
\label{sec:Nonper}

\subsection{The lowest partial wave}
\label{subsec:Lpw}

From the two-loop expression for the amplitude, Eq. (\ref{eq:disp2}),
one may project out the lowest $P$ partial wave with the use of
Eq. (\ref{eq:pwave}). This gives the expansion
\begin{equation}
\label{eq:ChPTlpar}
f_1(s)=f_1^{(0)}(s)+f_1^{(1)}(s)+f_1^{(2)}(s) ,
\end{equation}
which will satisfy the perturbative unitarity relations (\ref{eq:puni}).
These relations work very well at low energies, whereas the deviation
from exact unitarity, Eq. (\ref{eq:funi}), becomes more pronounced
as the energy is increased. At these energies, still higher order
unitarity corrections will start to be of importance in the chiral
expansion. This is particularly the case in the $\rho$(770) resonance
region, where unitarity corrections are essential.

In the presence of the $\rho$(770) resonance, unitarity will therefore
be of the utmost importance. There are different ways to
combine exact unitarity and the chiral expansion in order to try
to account for this resonance. One such method, which has been
successfully applied to many different processes, is the
so-called non-perturbative inverse amplitude method (IAM)
\cite{ref:Truong88,ref:DP93,ref:Han97,ref:OOP}. Since this is a
rather general method, it can also be straightforwardly applied
to the present case. The starting point for the IAM is to write down
a dispersion relation for the inverse of the partial wave $f_1$. In
this dispersion relation, exact unitarity and the chiral expansion
are used to compute the important right cut, whereas the left cut
and the subtraction constants are approximated by ChPT. If one
writes down a similar dispersion relation for the chiral expansion
using perturbative unitarity on the right cut, it is possible to
express the result of the IAM in a simple way in terms of the chiral
partial waves. With the two-loop ChPT expansion
(\ref{eq:ChPTlpar}), the result of the IAM can be written as
\begin{equation}
\label{eq:IAMpar}
f_1(s)=\frac{{f_1^{(0)}}^2(s)}{f_1^{(0)}(s)-f_1^{(1)}(s)+
{f_1^{(1)}}^2(s)/f_1^{(0)}(s)-f_1^{(2)}(s)} .
\end{equation}
This is formally equivalent to the [0,2] Pad\'{e} approximant
applied to ChPT and will therefore coincide with the chiral expansion
up to two loops. However, since exact unitarity was used in the
derivation of the IAM, it is expected that this result will improve
ChPT at higher energies. The IAM applied to two-loop ChPT has been
extensively discussed in Ref. \cite{ref:Han97}, where the detailed
derivation of the general result can also be found.

\begin{figure}[t]
\centerline{\psfig{figure=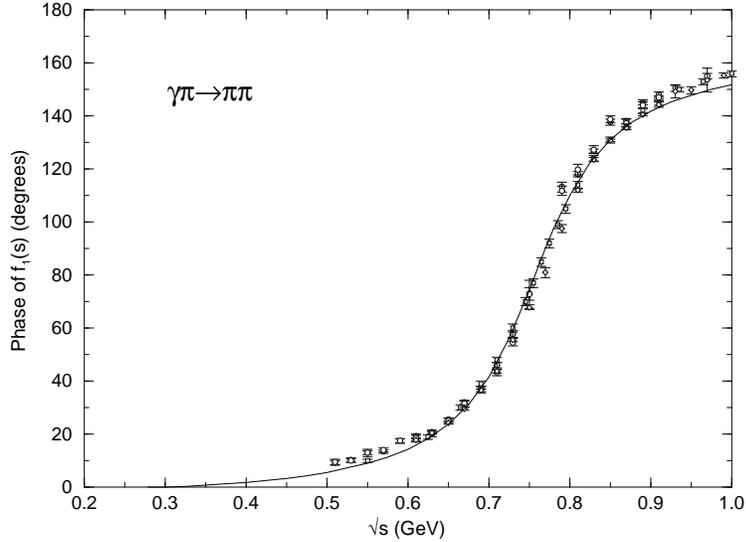,angle=-90}}
\caption{The phase of the lowest partial wave $f_1$ obtained
from the IAM, Eq. (\protect\ref{eq:IAMpar}), compared to
the experimental $\pi\pi$ phase shift $\delta_1^1$ from Ref.
\protect\cite{ref:EM74} (circles), Ref. \protect\cite{ref:Proto73}
(squares), and Ref. \protect\cite{ref:Hyams73} (diamonds).}
\label{fig1}
\end{figure}

\begin{figure}[t]
\centerline{\psfig{figure=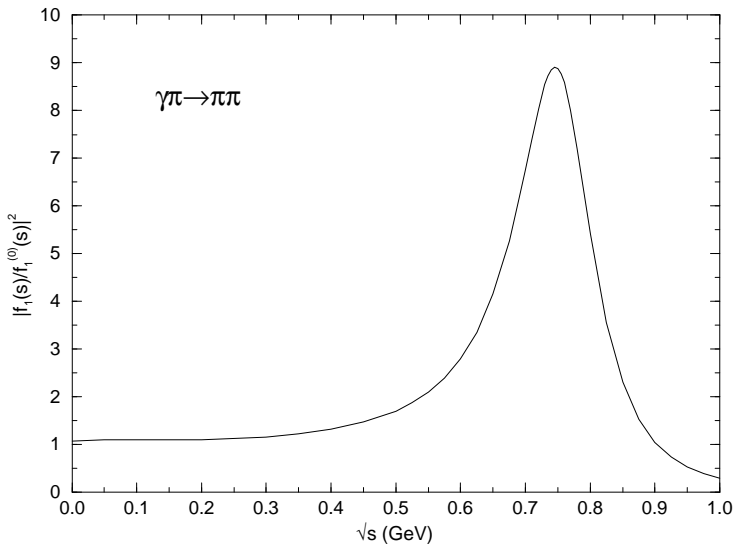,angle=-90}}
\caption{The lowest partial wave $|f_1/f_1^{(0)}|^2$ obtained
from the IAM, Eq. (\protect\ref{eq:IAMpar}).}
\label{fig2}
\end{figure}

The result of the IAM, Eq. (\ref{eq:IAMpar}), depends on the pion
mass and pion decay constant, which will be set equal to
$M_{\pi}=139.6$ MeV and $F_{\pi}=92.4$ MeV, respectively.
Furthermore, the IAM is also given in terms of the low-energy
constants $\bar{l}_2-\bar{l}_1$ and $\bar{l}_4$. These low-energy
constants appear in the ChPT $\pi\pi$ $P$ partial wave to one loop
order and can been determined phenomenologically from other sources.
However, since the IAM contains higher order unitarity corrections,
the phenomenological values of these low-energy constants in the IAM
do not necessarily coincide precisely with the values obtained in ChPT.
For the combination $\bar{l}_2-\bar{l}_1$, this has been determined
by applying the IAM to one-loop ChPT in the case of $\pi\pi$ scattering
with the result $\bar{l}_2-\bar{l}_1=5.8$ \cite{ref:Han97}. The
other low-energy constant $\bar{l}_4$ has been determined in ChPT
from the pion scalar form factor with the central value $\bar{l}_4=4.4$
\cite{ref:BCT98}. Since the IAM (\ref{eq:IAMpar}) does not depend
much on the precise value of this low-energy constant,
the same value can also be applied in the IAM. Finally, this result
also depends on the one-loop coupling constant $\bar{c}_1$. This
coupling constant is related to the low-energy constant $C_2^r$,
which has been estimated from the assumption of vector resonance
saturation, Eq. (\ref{eq:cr2}) \cite{ref:BBC90a}. With the value of
$C_2^r$ chosen at the renormalization scale $\mu =M_{\rho}$, one
finds $\bar{c}_1=1.71$. Therefore, in the IAM (\ref{eq:IAMpar}),
the values
\begin{equation}
\label{eq:IAMcon1}
\bar{l}_2-\bar{l}_1=5.8\;\;\;\; ,\;\;\;\;\bar{l}_4=4.4
\;\;\;\; ,\;\;\;\;\bar{c}_1 = 1.71
\end{equation}
will be used throughout. Having fixed these constants, it is possible
to determine the remaining two-loop coupling constants $\bar{c}_2$ and
$\bar{d}_2$ in the IAM from the mass and width of the $\rho$(770)
resonance. The mass of this resonance can be defined as the energy
where the phase passed $90^{\circ}$ and the width can be determined
from the slope of the phase at the resonance. With $M_{\rho}=770.0$
MeV and $\Gamma_{\rho}=150.7$ MeV \cite{ref:PDG98}, this gives the
values
\begin{equation}
\label{eq:IAMc2d2}
\bar{c}_2=-27\;\;\;\; ,\;\;\;\;\bar{d}_2=3.477 ,
\end{equation}
where the IAM depends rather strongly on $\bar{d}_2$ and to a lesser
extent on $\bar{c}_2$. With these values the IAM gives the phase
shown in Fig. \ref{fig1}. It is observed that the result agrees very
well with the experimental phase shifts all the way up to
1 GeV. Therefore, the IAM satisfies unitarity at least up to
this energy. In Fig. \ref{fig2}, the normalized absolute square of
the IAM partial wave is shown. In this case there is no
experimental data to be compared with. However, assuming that the  
full amplitude is completely dominated by the lowest partial wave
in the resonance region, the cross section (\ref{eq:totalcross})
can be expressed in terms of $|f_1|^2$. This cross section is usually
parameterized in the resonance region in terms of the
$\rho\rightarrow\pi\gamma$ partial width $\Gamma_{\pi\gamma}$
by using the Breit-Wigner form

\begin{equation}
\label{eq:BWsigma}
\sigma (s)=\frac{24\pi s}{(s-M_{\pi}^2)^2}\frac{M_{\rho}^2\Gamma_{\rho}
\Gamma_{\pi\gamma}}{(s-M_{\rho}^2)^2+M_{\rho}^2\Gamma_{\rho}^2} .
\end{equation}
Equating the two expressions for the cross section at the resonance
energy, the value $\Gamma_{\pi\gamma}=96$ KeV is obtained from
the IAM. Experimentally, the value of $\Gamma_{\pi\gamma}$ has
been extracted from the production of $\rho$ via the Primakoff
effect by incident pion on nuclear targets. The present
experimental data are, however, not very consistent with each other
\cite{ref:PDG98}. The latest experiment gave the value
$\Gamma_{\pi\gamma}=81\pm 4\pm 4$ KeV \cite{ref:Capra87}, whereas
two earlier experiments gave somewhat lower values for
$\Gamma_{\pi\gamma}$ \cite{ref:Huston86}. From the process
$e^+e^-\rightarrow\pi\gamma$, this partial width has also been
obtained giving the value $\Gamma_{\pi\gamma}=121\pm 31$ KeV
\cite{ref:Dol89}. Therefore, at present it can only be concluded that
the value for $\Gamma_{\pi\gamma}$ obtained from the IAM is not
in conflict with the experimental situation, but new experiments
in the resonance region are definitely needed.

\subsection{The full amplitude}
\label{subsec:Fuam}

If the absorptive part of the higher partial waves with $l\geq 3$
is neglected, it was shown in Sect. \ref{subsec:fispfor} that
the full amplitude could be constructed from the absorptive part
of the lowest partial wave. Since the IAM gives a good description
of this partial wave up to a least 1 GeV, one may use the
absorptive part of this partial wave in the Roy equation for
$\gamma\pi\rightarrow\pi\pi$, Eq. (\ref{eq:Ffull}). With the lowest
partial wave approximated by the IAM, the Roy equation may be written
with only one subtraction, which is fully determined by the chiral
anomaly result. Therefore, the non-perturbative chiral result for
the full amplitude can be written as
\begin{equation}
\label{eq:npcr}
F_{3\pi}(s,t,u) = F_{3\pi}^{(0)}+\frac{1}{\pi}
\int_{4M_{\pi}^2}^{\infty}ds'\frac{{\rm Im}f_1(s')}{s'}\left(
\frac{s}{s'-s}+\frac{t}{s'-t}+\frac{u}{s'-u}\right) ,
\end{equation}
where ${\rm Im}f_1$ is given by the absorptive part of the IAM result
(\ref{eq:IAMpar}). The full amplitude is by construction fully
symmetrical in $s$, $t$, and $u$. However, if the lowest partial wave
$f_1$ is projected out from Eq. (\ref{eq:npcr}), the result does not
agree exactly with the IAM result (\ref{eq:IAMpar}). This is due to the
fact that the left cut is now fully determined from crossing symmetry
instead of being approximated by the chiral expansion. In the
elastic region below 1 GeV, the difference is, however, negligible.
From the full amplitude, Eq. (\ref{eq:npcr}), it is also found
that the higher partial waves are very small below 1 GeV. Therefore,
the cross section can indeed be expressed in terms of $|f_1|^2$ to
a very good approximation. For a somewhat different evaluation of
the Roy equation for $\gamma\pi\rightarrow\pi\pi$, see
Ref. \cite{ref:Truong99}.

As the IAM was derived using elastic unitarity, it is expected
that this result is only applicable in the elastic region.
The high energy part of the
dispersion relation (\ref{eq:npcr}) is therefore not expected to
be very well approximated by the IAM. Still, at low and moderate
energies the dispersion relation should be almost completely
saturated by the $\rho$(770) contribution. Therefore, at these
energies the high energy part of the dispersion relation
should not be very important. Furthermore, at these energies
it is also expected that the contribution from the higher
partial waves to the Roy equation should be negligible.
Therefore, the non-perturbative chiral result (\ref{eq:npcr})
should give a rather accurate description of the full amplitude
at low and moderate energies. From this result, it is possible to
determine the subtraction constants $\bar{C}$ and $\bar{D}$
in the two-loop ChPT result (\ref{eq:disp2}) via the sum rules
\begin{equation}
\label{eq:sumrules}
\bar{C} = \frac{1}{F_{3\pi}^{(0)}}\frac{1}{\pi}\int_{4M_{\pi}^2}^
{\infty}ds'\frac{{\rm Im}f_1(s')}{s'^2}\;\;\;\; ,\;\;\;\;
\bar{D} = \frac{1}{F_{3\pi}^{(0)}}\frac{1}{\pi}\int_{4M_{\pi}^2}^
{\infty}ds'\frac{{\rm Im}f_1(s')}{s'^3} .
\end{equation}
Evaluating these sum rules, one obtains the values $\bar{C}=0.93$
${\rm GeV}^{-2}$ and $\bar{D}=2.12$ ${\rm GeV}^{-4}$, respectively.
These sum rules are indeed almost completely saturated
by the dispersion relation up to 1 GeV, as the higher energy part
only contributes by approximately 2\% for $\bar{C}$ and approximately
1\% for $\bar{D}$. With the one-loop coupling constant $\bar{c}_1$
determined from the assumption of vector resonance saturation,
Eq. (\ref{eq:IAMcon1}), this gives the following values for the
two-loop coupling constants
\begin{equation}
\label{eq:ChPTc2d2}
\bar{c}_2=-20\;\;\;\; ,\;\;\;\;\bar{d}_2=2.7 .
\end{equation}
These values are slightly different from the values determined
directly from the IAM (\ref{eq:IAMc2d2}). As it has already been
stated, this is indeed to be expected since the IAM contains higher
order unitarity effects, which will effect the determination of
these coupling constants.

Another way to combine unitarity and the chiral expansion has
been proposed by Holstein \cite{ref:Holstein96}. In this model,
vector meson dominance is combined with one-loop ChPT in order
to account for rescattering effects. Projecting out the lowest
partial wave, it is found that unitarity is
approximately satisfied and that the $\rho\rightarrow\pi\gamma$
partial width is given by $\Gamma_{\pi\gamma}=84$ KeV. Within
this model, one obtains the values
\begin{equation}
\label{eq:holsteincon}
\bar{c}_2=-19\;\;\;\; ,\;\;\;\;\bar{d}_2=2.9 ,
\end{equation}
which agree very well with the determination given in
(\ref{eq:ChPTc2d2}). However, from the model proposed by Holstein,
all higher partial waves have the same phase as the lowest
partial wave, which is in contradiction with the experimental
facts. In order to remedy this, the lowest partial wave obtained
in this model may be inserted in the dispersion relation
(\ref{eq:npcr}). The resulting amplitude still agrees well
with the experimental information, but now implies that all the
higher partial waves are real. This gives the values $\bar{c}_2=-16$
and $\bar{d}_2=2.8$ from the sum rules (\ref{eq:sumrules}),
which is also fully consistent with the values given in
(\ref{eq:ChPTc2d2}). Therefore, it seems that the
solution of the sum rules is rather robust against small
variations of the input.

In the numerical results for two-loop ChPT (\ref{eq:disp2}), the
values of $\bar{c}_2$ and $\bar{d}_2$ determined in
(\ref{eq:ChPTc2d2}) will be used together with the value of
$\bar{c}_1$ given in (\ref{eq:IAMcon1}). For the low-energy
constants $\bar{l}_1$ and $\bar{l}_2$, these have been determined
in ChPT. With the use of the two-loop result for the $\pi\pi$
scattering amplitude and fitting the $D$ partial wave scattering
lengths, the value $\bar{l}_2-\bar{l}_1=6.0$ has been obtained
\cite{ref:Bijetal96}. Similar values have also recently been obtained
from a two-loop calculation of the $K_{l4}$ form factors
\cite{ref:ABT99}. These values are rather consistent with the value
given in (\ref{eq:IAMcon1}) but are somewhat lower than the value
$\bar{l}_2-\bar{l}_1=7.8$ obtained from a dispersive improved
one-loop calculation of the $K_{l4}$ form factors \cite{ref:BCG94}.
However, the numerical result only depends very slightly on the
exact value of $\bar{l}_2-\bar{l}_1$. Therefore, the values given in
Eq. (\ref{eq:IAMcon1}) for both $\bar{l}_2-\bar{l}_1$ and
$\bar{l}_4$ will also be used for ChPT.

\section{Numerical results}
\label{sec:Num}

\subsection{Amplitude and cross section}
\label{subsec:Amp}

\begin{figure}[t]
\centerline{\psfig{figure=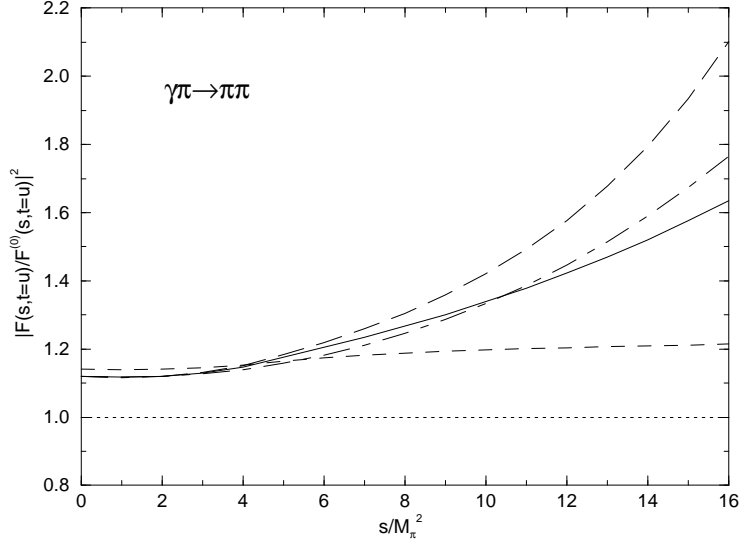,angle=-90}}
\caption{The amplitude $|F_{3\pi}/F_{3\pi}^{(0)}|^2$ evaluated
at $t=u$ as a function of $s/M_{\pi}^2$. The solid line is the
two-loop result, Eq. (\protect\ref{eq:disp2}), the dashed line
the one-loop result, Eq. (\protect\ref{eq:ChPT1}), and the
dotted line the leading order chiral anomaly result,
Eq. (\protect\ref{eq:ChPT0}). Finally, the long-dashed line is
the non-perturbative chiral result, Eq. (\protect\ref{eq:npcr}),
and the dash-dotted line is the contribution from the subtraction
constants.}
\label{fig3}
\end{figure}

Fig. \ref{fig3} shows the absolute square of the amplitude normalized
to the chiral anomaly. In the sub-threshold region, the two-loop result
is rather close to the one-loop result, whereas the corrections start
to be of importance slightly above threshold. As for the one-loop
result, this gives an almost constant correction to the leading order
chiral anomaly result. This is due to the fact that the only variation
comes from the $s$, $t$, and $u$ dependency of the function $F$ in the
one-loop expression (\ref{eq:ChPT1}). The contribution from the
subtraction constants is also shown in the figure, where this
contribution is obtained from the expression
$F_{3\pi}(s,t,u)=F_{3\pi}^{(0)}[1+\bar{C}(s+t+u)+\bar{D}(s^2+t^2+u^2)]$.
It is observed that this part gives the main
contribution to the two-loop
result, which is due to the presence of the $\rho$(770) resonance.
However, the unita\-rity corrections coming from the $U(x)$ terms
are also of some importance in the numerical result. In fact, these
unitarity corrections are essential in the non-perturbative chiral
result, Eq. (\ref{eq:npcr}), which is also shown in the figure.
The deviation between this result and the two-loop chiral result
gives an estimate of the range of validity of the truncated chiral
expansion. From this deviation, it is observed
that still higher order chiral corrections
should begin to be of some importance somewhat around $s=10M_{\pi}^2$,
which is also observed in other two-loop calculations.

\begin{figure}[t]
\centerline{\psfig{figure=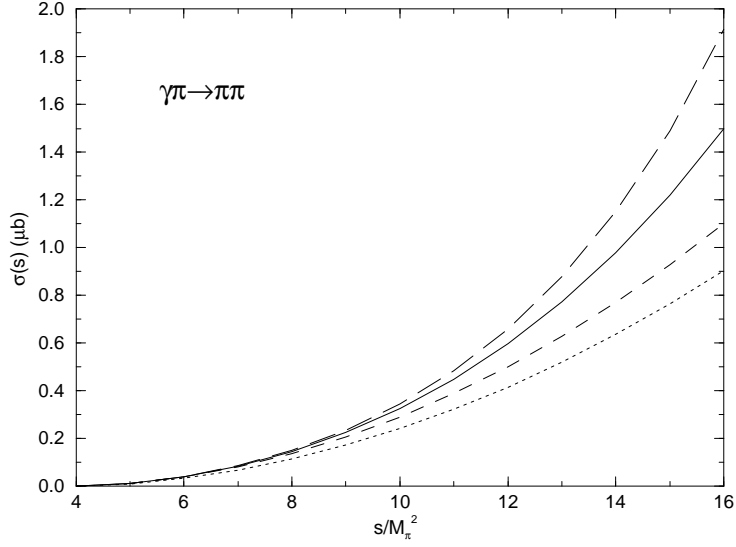,angle=-90}}
\caption{The total cross section as a function of $s/M_{\pi}^2$. The
solid line is the two-loop result, the dashed line the one-loop
result, the dotted line the leading order result and the long-dashed
line the non-perturbative chiral result.}
\label{fig4}
\end{figure}

In Fig. \ref{fig4}, the total cross section is shown as a function
of $s/M_{\pi}^2$, where the total cross section is given by
Eq. (\ref{eq:totalcross}). Since the one-loop expression for
$F_{3\pi}(s,t,u)$ is more or less constant in the whole phase space,
the one-loop result for the cross section gives an almost constant
correction to the leading order result of approximately 20 \%. Close
to threshold, the additional two-loop correction is rather small
compared to the one-loop correction, whereas this is not the
case for higher energies. The additional two-loop correction amounts to
approximately 75 \% of the one-loop correction at $s=10M_{\pi}^2$, and
even more at higher energies. This indicates that even higher order
terms in the chiral expansion should begin to be of importance at
this energy. This is also observed in the figure, where the
non-perturbative chiral result begins to deviate from the two-loop
result around this energy.

\subsection{Comparison with experiments}
\label{subsec:Exp}

The process $\gamma\pi\rightarrow\pi\pi$ has been investigated
at low energies by the Serpukhov experiment \cite{ref:Antipov97}.
This experiment used the Primakoff reaction of pion pair production
by pions in the nuclear Coulomb field
\begin{equation}
\label{eq:primakoff}
\pi^-+(Z,A)\rightarrow\pi^-+\pi^0+(Z,A) .
\end{equation}
The cross section for this process is related to the
$\gamma\pi\rightarrow\pi\pi$ cross section through the
equivalent photon method
\begin{equation}
\label{eq:epm}
\frac{d\sigma}{dsdtdq^2} = \frac{Z^2\alpha}{\pi}\left[
\frac{q^2-q^2_{\rm min}}{q^4}\right] \frac{1}{s-M^2_{\pi}}
\frac{d\sigma_{\gamma\pi\rightarrow\pi\pi}}{dt} ,
\end{equation}
where
\begin{equation}
\label{eq:gpcross}
\frac{d\sigma_{\gamma\pi\rightarrow\pi\pi}}{dt} =
\frac{|F_{3\pi}(s,t,u)|^2}{512\pi}(s-4M^2_{\pi})\sin^2\theta
\end{equation}
and
\begin{equation}
\label{eq:q2min}
q^2_{\rm min} = \left(\frac{s-M^2_{\pi}}{2E}\right)^2
\end{equation}
and $E$ is the energy of the incident pion beam. Neglecting
the $q^2$ dependency in $F_{3\pi}(s,t,u)$, i.e., setting
$q^2\approx 0$ so that $s+t+u=3M_{\pi}^2$, the total cross
section is given by
\begin{eqnarray}
\label{eq:intcross}
\sigma & = & \frac{Z^2\alpha}{1024\pi^2}\int_{4M^2_{\pi}}^
{s_{\rm max}}ds\frac{(s-4M^2_{\pi})^{3/2}}{s^{1/2}}\left[ \ln
\frac{q^2_{\rm max}}{q^2_{\rm min}}+\frac{q^2_{\rm min}}
{q^2_{\rm max}}-1\right]
\times \nonumber \\
&& \int_0^{\pi}d\theta\sin^3\theta |F_{3\pi}(s,t,u)|^2 .
\end{eqnarray}
The Serpukhov experiment was carried out with a 40 GeV pion
beam and the maximum momentum transfer was
$q^2_{\rm max}=2\times 10^{-3}$ ${\rm GeV}^2\ll M_{\pi}^2$.
Thus, the $q^2$ dependency in $F_{3\pi}(s,t,u)$ can
indeed be neglected when one compares with this experiment.

The kinematical region studied in the Serpukhov experiment
was restricted to $s_{\rm max}=10M_{\pi}^2$. Three different
targets (C, Al, and Fe) were used and the measured total cross
sections were found to agree well with the theoretical $Z^2$
dependence. Averaging the values of $\sigma /Z^2$ obtained
from the three different targets, the experimental result
given in Table \ref{table1} was obtained. This experimental
result has to be compared with the theoretical predictions, which are
also given in Table \ref{table1}. It is observed that the
two-loop contribution gives a sizable correction to the
one-loop result, whereas the further correction from the
non-perturbative chiral result is rather small. Nevertheless, all
results are still somewhat below the central experimental value,
which, however, also has rather large error bars.

\begin{table}[t]
\caption{The total cross section for the Primakoff reaction
(\ref{eq:primakoff}) divided by $Z^2$. The theoretical predictions
are obtained from Eq. (\protect\ref{eq:intcross}) with
$F_{3\pi}(s,t,u)$ given by the leading order chiral anomaly result
($O(p^4)$), the one-loop result ($O(p^6)$), the two-loop result
($O(p^8)$), and the non-perturbative chiral result (NPCR). The
experimental result is the average value obtained in Ref.
\protect\cite{ref:Antipov97}, where the first (second) error is
statistical (systematic).}
\label{table1}
\vspace{0.2cm}
\begin{center}
\begin{tabular}{cccccc}
\hline
&$O(p^4)$&$O(p^6)$&$O(p^8)$&NPCR&Experiment\\
\hline
$\sigma /Z^2$ (nb)&0.92&1.09&1.18&1.21&$1.63\pm 0.23\pm 0.13$\\
\hline
\end{tabular}
\end{center}
\end{table}

\begin{table}[t]
\caption{The chiral anomaly $F_{3\pi}^{(0)}$ extracted from the
experimental data. The different va\-lues correspond to setting
$F_{3\pi}(s,t,u)$ in Eq. (\protect\ref{eq:intcross})
equal to the leading order chiral anomaly result ($O(p^4)$),
the one-loop result ($O(p^6)$), the two-loop result ($O(p^8)$),
and the non-perturbative chiral result (NPCR). The theoretical
result is from Eq. (\protect\ref{eq:ChPT0}) with
$F_{\pi}=92.4\pm 0.3$ MeV \protect\cite{ref:PDG98}.}
\label{table2}
\vspace{0.2cm}
\begin{center}
\begin{tabular}{cccccc}
\hline
&$O(p^4)$&$O(p^6)$&$O(p^8)$&NPCR&Theory\\
\hline
$F_{3\pi}^{(0)}$ (${\rm GeV}^{-3}$)&
$12.9\pm 1.4$&$11.9\pm 1.3$&$11.4\pm 1.3$&
$11.3\pm 1.3$&$9.7\pm 0.1$\\
\hline
\end{tabular}
\end{center}
\end{table}

From the experimental result of $\sigma /Z^2$, it is
possible to extract the value of the chiral anomaly
$F_{3\pi}^{(0)}$, if this is regarded as a free parameter.
These extracted values are shown in Table \ref{table2},
where the statistical and systematic errors have been added.
The theoretical prediction obtained from Eq. (\ref{eq:ChPT0})
is also shown in this table. With $F_{3\pi}(s,t,u)$
given by the leading order chiral anomaly result, one
obtains a value of $F_{3\pi}^{(0)}$ which is 2.3$\sigma$
too high compared with the theoretical prediction.
This result for $F_{3\pi}^{(0)}$ is the value generally
quoted from the analysis in Ref. \cite{ref:Antipov97}.
However, with the one-loop expression for $F_{3\pi}(s,t,u)$,
one obtains a value of $F_{3\pi}^{(0)}$ which is only
1.7$\sigma$ too high, and with the two-loop result the
disagreement is only at the 1.3$\sigma$ level.

Since the non-perturbative chiral result is close to the two-loop
result, the uncertainty coming from yet higher orders
in the chiral expansion should be rather small, at 
least in the kinematical region probed by the Serpukhov
experiment. There is of course also some uncertainty due to
the uncertainty in the determination of the coupling constants
$\bar{c}_2$ and $\bar{d}_2$. However, as already discussed, this
uncertainty should also be rather small. Therefore, in view of the
disagreement between the theoretical predictions and the Serpukhov
experiment, new improved Primakoff experiments are definitely needed.

The process $\gamma\pi\rightarrow\pi\pi$ has also been
investigated at low energies at CERN \cite{ref:Amen85}. This
experiment used the reaction $\pi^-e\rightarrow\pi^-\pi^0e$
with 300 GeV pions and measured the total cross section
for this process. It was found that the cross section obtained
agreed with the chiral anomaly prediction with three
number of colors. However, since the error bars were rather
large, it was not possible to observe any systematic deviations
from the soft pion and soft photon limit in this experiment.

Therefore, new precision experiments are necessary in order
to investigate the process $\gamma\pi\rightarrow\pi\pi$ and
thus the theory of chiral anomalies in greater detail.
Indeed, such new experiments are under way at different
facilities. In the COMPASS experiment at CERN \cite{ref:COMPASS}
and in the SELEX experiment at FNAL (E781) \cite{ref:SELEX},
the Primakoff reaction (\ref{eq:primakoff}) will be measured
with 600 and 50-280 GeV pion beams, respectively. In these two
experiments, the expected number of near threshold two-pion
events is several orders of magnitude higher than previously
obtained. This will allow analysis of the data separately
in different intervals of $s$ with small statistical
errors.

The SELEX experiment will also measure the reaction
$\pi^-e\rightarrow\pi^-\pi^0e$ in order to obtain
independent information on the $\gamma\pi\rightarrow\pi\pi$
amplitude. For this reaction, the expected number of events
is also significantly larger than previously obtained,
which would give excellent complementary information
on $F_{3\pi}(s,t,u)$.

Finally, at CEBAF the process
$\gamma\pi\rightarrow\pi\pi$ is investigated
by measuring $\gamma p\rightarrow\pi^+\pi^0 n$ cross sections
using tagged photons \cite{ref:CEBAF}. Since the resonance
region will also be measured in this experiment, this
could give new information on the $\rho\rightarrow\pi\gamma$
partial width $\Gamma_{\pi\gamma}$. However, as the
incident pion is virtual in the CEBAF experiment, this has
to be taken into account when comparing with theory.

\section{Conclusion}
\label{sec:Con}

The anomalous process $\gamma\pi\rightarrow\pi\pi$ plays an important
role in the theory of chiral anomalies. At leading order in the
chiral expansion, which corresponds to the soft pion and soft photon
limit, the amplitude for this process is given in terms of the number
of colors $N_c$ of the strong interaction. Comparing the leading order
result with the present experimental information \cite{ref:Antipov97},
one finds that the value $N_c=4$ is favored.

In order to test this important conclusion more precisely, it is
necessary with both experimental and theoretical improvements. Indeed,
significant improvements on the experimental side are expected from
several new experiments \cite{ref:COMPASS,ref:SELEX,ref:CEBAF}. On the
theoretical side, the one-loop correction to the leading order result
has previously been calculated \cite{ref:BBC90a}. However, in view of
the new precision experiments, it is important also to calculate the
additional two-loop correction to the amplitude.

This has been the purpose of the present paper, where the amplitude
for the anomalous process $\gamma\pi\rightarrow\pi\pi$ has been
evaluated to two loops in the chiral expansion by means of a dispersive
method. The two new coupling constants that enter at two-loop order
were determined from sum rules with the use of a non-perturbative
chiral approach. The uncertainty in the numerical results due to this
determination was estimated to be rather small. Moreover, the still
higher order terms in the chiral expansion were also estimated to
be small at low energies.

The two-loop result improves the agreement
with the present experimental information \cite{ref:Antipov97}
compared to both the leading order and the one-loop results.
However, the two-loop prediction is still significantly below
the central experimental data for $N_c=3$. This fact
is not likely to be due to theoretical uncertainties. Therefore,
should the new experiments \cite{ref:COMPASS,ref:SELEX,ref:CEBAF}
confirm the present central experimental value with better precision,
it would be a serious problem for QCD.

\ack

The author thanks T. N. Truong for useful discussions.
Part of this work was done while the author was visiting Ecole
Polytechnique, Paris, which is acknowledged for its hospitality
and financial support.


\begin{thebibliography}{99}
\bibitem{ref:We79} S. Weinberg, Physica A 96 (1979) 327.
\bibitem{ref:GL84} J. Gasser and H. Leutwyler, Ann. Phys. (N.Y.)
158 (1984) 142.
\bibitem{ref:GL85} J. Gasser and H. Leutwyler, Nucl. Phys. B 250 (1985)
465.
\bibitem{ref:Leut94} H. Leutwyler, Ann. Phys. (N.Y.) 235 (1994) 165.
\bibitem{ref:RevChPT} For some reviews on ChPT see e.g.
U. G. Mei{\ss}ner, Rep. Prog. Phys. 56 (1993) 903;
A. Pich, Rep. Prog. Phys. 58 (1995) 563;
G. Ecker, Prog. Part. Nucl. Phys. 35 (1995) 1;
J. Bijnens and U. G. Mei{\ss}ner, Proc. Workshop on Chiral Effective
Theories, Bad Honnef, Germany, 30 November - 4 December 1998
(hep-ph/9901381).
\bibitem{ref:GM91} J. Gasser and U. G. Mei{\ss}ner, Nucl. Phys. B 357
(1991) 90; G. Colangelo, M. Finkemeier and R. Urech, Phys. Rev. D 54
(1996) 4403.
\bibitem{ref:BCT98} J. Bijnens, G. Colangelo and P. Talavera,
JHEP 05 (1998) 014.
\bibitem{ref:Knecht95} M. Knecht {\em et al.}, Nucl. Phys. B 457
(1995) 513.
\bibitem{ref:Bijetal96} J. Bijnens {\em et al.}, Phys. Lett. B 374
(1996) 210; Nucl. Phys. B 508 (1997) 263; 517 (1998) 639 (E).
\bibitem{ref:BGS94} S. Bellucci, J. Gasser and M. E. Sainio, Nucl.
Phys. B 423 (1994) 80; B 431 (1994) 413 (E);
U. B\"{u}rgi, Phys. Lett. B 377 (1996) 147; Nucl. Phys. B 479 (1996)
392;
J. Bijnens and P. Talavera, Nucl. Phys. B 489 (1997) 387.
\bibitem{ref:WZ71} J. Wess and B. Zumino, Phys. Lett. 37 B (1971) 95;
E. Witten, Nucl. Phys. B 223 (1983) 422.
\bibitem{ref:BBC90} J. Bijnens, A. Bramon and F. Cornet, Z. Phys.
C 46 (1990) 599;
D. Issler, preprint SLAC-PUB-4943;
R. Akhoury and A. Alfakih, Ann. Phys. (N.Y.) 210 (1991) 81;
H. W. Fearing and S. Scherer, Phys. Rev. D 53 (1996) 315.
\bibitem{ref:Bij93} For a review on ChPT in the anomalous sector
see J. Bijnens, Int. J. Mod. Phys. A 8 (1993) 3045.
\bibitem{ref:Adler71} S. L. Adler {\em et al.}, Phys. Rev. D 4 (1971)
3497;
M. V. Terent'ev, Phys. Lett. 38 B (1972) 419;
R. Aviv and A. Zee, Phys. Rev. D 5 (1972) 2372.
\bibitem{ref:BBC90a} J. Bijnens, A. Bramon and F. Cornet, Phys. Lett.
B 237 (1990) 488.
\bibitem{ref:Antipov97} Yu. M. Antipov {\em et al.}, Phys. Rev. D 36
(1987) 21.
\bibitem{ref:COMPASS} M. A. Moinester, V. Steiner and S. Prakhov,
{\em in} Proc. XXXVII Int. Winter Meeting on Nuclear Physics, Bormio,
Italy, 25 - 29 January 1999, ed. I. Iori (Ricerca Scientifica ed
Educazione Permanente Supplemento No. 114, 1999), preprint
TAUP-2562-99 (hep-ex/9903017).
\bibitem{ref:SELEX} SELEX Collaboration, M. A. Moinester {\em et al.},
{\em in} Proc. 8th Int. Conf. on the Structure of Baryons
(BARYONS '98), Bonn, Germany, 22 - 26 September 1998,
eds. D. W. Menze and B. Metsch (World Scientific, Singapore, 1999),
preprint TAUP-2568-99 (hep-ex/9903039); M. A. Moinester,
{\em in} Proc. Int. Conf. on Physics with GeV Particle Beams,
J\"{u}lich, Germany, 22 - 25 August 1994, eds. H. Machner and K.
Sistemich (World Scientific, N.J., 1994), preprint TAUP-2176-94
(hep-ph/9409307).
\bibitem{ref:CEBAF} R. A. Miskimen, K. Wang and A. Yegneswaran,
Spokesmen, CEBAF Proposal PR-94-015, 1994.
\bibitem{ref:Watson54} K. M. Watson, Phys. Rev. 95 (1954) 228.
\bibitem{ref:Adler69} S. L. Adler, Phys. Rev. 177 (1969) 2426;
J. S. Bell and R. Jackiw, Nuovo Cimento A 60 (1969) 47;
W. A. Bardeen, Phys. Rev. 184 (1969) 1848.
\bibitem{ref:DHL85} J. F. Donoghue, B. R. Holstein and Y. C. R. Lin,
Phys. Rev. Lett. 55 (1985) 2766; J. Bijnens, A. Bramon and F. Cornet,
Phys. Rev. Lett. 61 (1988) 1453.
\bibitem{ref:Ecker89} G. Ecker {\em et al.}, Nucl. Phys. B 321 (1989)
311; G. Ecker {\em et al.}, Phys. Lett. B 223 (1989) 425.
\bibitem{ref:Roy71} S. M. Roy, Phys. Lett. 36 B (1971) 353;
J. L. Basdevant, J. C. Le Guillou and H. Navelet, Nuovo Cimento 7 A
(1972) 363; J. L. Petersen, CERN Yellow Report, CERN 77-04.
\bibitem{ref:SSF93} J. Stern, H. Sazdjian and N. H. Fuchs, Phys. Rev. D
47 (1993) 3814.
\bibitem{ref:Truong88} T. N. Truong, Phys. Rev. Lett. 61 (1988) 2526;
67 (1991) 2260.
\bibitem{ref:DP93} A. Dobado and J. R. Pel\'{a}ez, Phys. Rev. D 47
(1993) 4883; 56 (1997) 3057; T. Hannah, Phys. Rev. D 51 (1995) 103;
52 (1995) 4971; 54 (1996) 4648.
\bibitem{ref:Han97} T. Hannah, Phys. Rev. D 55 (1997) 5613.
\bibitem{ref:OOP} J. A. Oller, E. Oset and J. R. Pel\'{a}ez,
Phys. Rev. Lett. 80 (1998) 3452; Phys. Rev. D 59 (1999) 074001;
60 (1999) 099906 (E); F. Guerrero and J. A. Oller, Nucl. Phys. B 537
(1999) 459.
\bibitem{ref:PDG98} Particle Data Group, C. Caso {\em et al.},
Eur. Phys. J. C 3 (1998) 1.
\bibitem{ref:EM74} P. Estabrooks and A. D. Martin, Nucl. Phys. B 79
(1974) 301.
\bibitem{ref:Proto73} S. D. Protopopescu {\em et al.}, Phys. Rev. D 7
(1973) 1279.
\bibitem{ref:Hyams73} B. Hyams {\em et al.}, Nucl. Phys. B 64 (1973)
134; W. Ochs, Ph.D. thesis, Ludwig-Maximilians-Universit\"{a}t, 1973.
\bibitem{ref:Capra87} L. Capraro {\em et al.}, Nucl. Phys. B 288
(1987) 659.
\bibitem{ref:Huston86} J. Huston {\em et al.}, Phys. Rev. D 33 (1986)
3199; T. Jensen {\em et al.}, Phys. Rev. D 27 (1983) 26.
\bibitem{ref:Dol89} S. I. Dolinsky {\em et al.}, Z. Phys. C 42 (1989)
511.
\bibitem{ref:Truong99} T. N. Truong, {\em in} Proc. 4th Workshop
on Quantum Chromodynamics, Paris, France, 1 - 6 June 1998,
eds. H. M. Fried and B. Mueller (World Scientific, Singapore, 1999)
(hep-ph/9903378).
\bibitem{ref:Holstein96} B. R. Holstein, Phys. Rev. D 53 (1996) 4099.
\bibitem{ref:ABT99} G. Amor\'{o}s, J. Bijnens and P. Talavera,
Phys. Lett. B 480 (2000) 71; Nucl. Phys. B 585 (2000) 293.
\bibitem{ref:BCG94} J. Bijnens, G. Colangelo and J. Gasser,
Nucl. Phys. B427 (1994) 427.
\bibitem{ref:Amen85} S. R. Amendolia {\em et al.}, Phys. Lett. 155 B
(1985) 457.
\end{thebibliography}
\end{document}